\documentclass[english]{article}
\usepackage[T1]{fontenc}
\usepackage[latin1]{inputenc}
\usepackage{floatflt}
\usepackage{graphicx}

\makeatletter

 \newenvironment{lyxcode}
   {\begin{list}{}{
     \setlength{\rightmargin}{\leftmargin}
     \setlength{\listparindent}{0pt}
     \raggedright
     \setlength{\itemsep}{0pt}
     \setlength{\parsep}{0pt}
     \normalfont\ttfamily}%
    \item[]}
   {\end{list}}

\usepackage{babel}
\makeatother
\begin{document}

\title{A Model with a Cosmographic Landscape}

\maketitle
\begin{lyxcode}

\begin{center}\textrm{\textsc{Fritz~W.~Bopp}}\\
\textrm{\textsc{University~of~Siegen}}\end{center}
\end{lyxcode}
\begin{abstract}
To argue against a too narrow focus in the LHC Higgs search, a simple-minded
model with a rich ``cosmographic'' vacuum structure for the generation
of masses is developed on a conceptual level. In this framework Higgs
like bosons which could exist in the LHC mass range have no preference
to decay in heavy flavors. 
\end{abstract}

\section{Introduction}

Most theoretical concepts of LHC physics concerning the origin of
masses were developed three decades ago. It is important to keep in
mind that in spite of this respectable age there were few real tests
and our understanding of how masses arise still is in an early state.
In absence of experiments one had to search for theoretically convincing
theories. As experimental information from LHC is imminent it might
be wise to somewhat disrespectfully bet on simple-minded more generic
considerations. 

Fermion masses might reflect their coupling to a Higgs field also
responsible for the $W$ mass or not. The first possibility, the standard
model case, is carefully studied. For the second options there are
again two possibilities: fermions might not couple at all to the $W$-mass
Higgs field or fermions might couple to Higgs fields, but not proportional
to their mass. The first case of this option is again carefully considered
as it happens to correspond to the situation of the minimal super-symmetric
Higgs. Our focus here will be the second case.

As there is presumably less suppression of light fermions, its experimental
signature will be a pair of possible multi-TeV jets. It was considered
as a possible signal for something outside of the standard model at
Fermilab until the structure functions were better understood. Without
encouragement from theory dijets are still studied \cite{Fermilab}. 

To motivate such a search we outline a model where such a channel
would be central. In section 2 the basic concept is developed and
the fermion mass matrices are considered. The boson masses are the
topic of section 3. A discussion of possible experimental observation
and tests follows in section 4.

\section{The Model for the Fermion Masses}

For the motivation of the novel concept accept a fantasy. \textit{Imagine
Newton had been rewarded with really appropriate funding: The apple
tree had turned into ballistic constructions. Skills about very precise
triangulation and methods to correct for air friction might have been
developed and a final experiment with a huge cannon shooting from
coast to coast at an empty place in the new North American territories
might have led to a seemingly new physics result $V\sim1/r^{1+\epsilon}$.
A closer inspection would have led to the IVth Newton principle:}
\textbf{\textit{\emph{}}}\textit{NEVER MIX UP PURE PHYSICS WITH GEOGRAPHIC
ACCIDENTS -- in the fantasy the Rocky Mountains and the Appalachians.} 

The idea is that some of today's multitude of ``physical'' constants
cannot be calculated from first principles of physics but have to
be attributed to a basically accidental structure of a cosmographic
vacuum in our zone of the universe. 

This basic philosophy is of course not new. Similar ideas were, p.e.,
formulated in a more sophisticated framework of a Multiverse \cite{Weinberg,Wilczek},
which are based on the anthropic principle and which (usually) assume
cosmological uniformity. Even in the standard model the vacuum is
somewhat messy. It is known to contain a certain amount of chiral
condensates and one Higgs field with a complicated potential. The
separation of U(1) and SU(2) is not respected. So it seems not very
bold to accept a richer vacuum structure.

The term ``vacuum'' might be confusing. In field theory the vacuum
has no time and space dependence. The cosmological vacuum structure
locally mimics the field theoretical vacuum to which it corresponds
in the limit of a stable infinite extension of the states it contains.
The locality concept is not far fetched. It is usually assumed that
somehow the vacuum has a time dependence on a cosmic scale and relativistically
a space dependence is therefore in some way anyhow included.

The cosmographic vacuum can contain a largely accidental mixture of
fermion condensates. The effective negative binding energy is of the
order of the masses bound and the following condensation relation
\[
V_{\mathrm{{binding}}}+\sum m_{\mathrm{{fermion}}}+E_{\mathrm{{localization}}}\approx0\]
 can approximately be satisfied. A residual fermionic repulsion limits
the vacuum density. In this way a collapse in an infinite dense vacuum
is not possible. Fermions are usually not neutral. The absence of
interactions with $\gamma$ requires that these composite states of
the charged fermions are bound very tightly. 

The usual spontaneously broken vacuum leads to a problem. The energy
shifts during phase transitions have to be of the scale of particle
physics many orders of magnitude above the energy density needed in
cosmology. The cosmographic vacuum offers a solution of this hierarchy
problem. After condensing the cosmological vacuum preceded on a path
reducing its energy from a particle scale to a value close to zero.
The central assumption is that there is no intrinsic scale in this
decay. This $\partial\mu_{vac.}/\partial(\mu_{vac.}t)=-\kappa\mu_{vac.}$
leads to a simple power decrease, i.e. $\mu_{vac.}=\frac{1}{\kappa}/(t-t_{0})$.
Here $\mu_{vac.}$ is the energy density of the vacuum and $\kappa$
a dimensionless decay constant. The present value depends on the time
it had i.e.\ on the apparent age of the universe. As needed the dark
energy has a cosmological and not a particle physics scale.

This has an immediate consequence for the extension of the condensate
i.e.\ of $\left<<(\sum r_{i})^{2}>\sim\mathrm{{[cosmological\, scale]}}\right>$.
At the starting point of the condensation, i.e.\ at the phase-transition
temperature, it must have had a particle physics scale to overlap
with the visible world; nowadays it must be extremely extended reflecting
the age of the universe. Any decay and any change must now be extremely
slow. In this way consistency is reached with the observed homogeneity
of the physical laws which depend on the cosmographic vacuum.

In the standard model all parameters of the mass matrices are transferred
to coupling constants:\[
m_{ij}=g{}_{ij}\left<h\right>\]
 The basic idea of the cosmographic vacuum involves a simple change.
The couplings are essentially part of pure physics and unique. The
multitude of other parameters is simply transferred to properties
of richly structured, largely accidental cosmographic vacuum: \[
m_{ij}=\widetilde{g}\left<h{}_{ij}\right>\]
In the considered realization of the cosmographic vacuum the Higgs
fields $h{}_{ij}$ correspond to fermion condensates.

The non-empty vacuum does select a Lorentz system. The observed (effective)
Lorentz invariance of the visible world limits the interaction with
the cosmographic vacuum to a purely Lorentz scalar one. The physical
mechanism -- how this works, lies in a low energy effective theory
based on the the Operator Product Expansion. Tensor coupling with
non-zero rank involves derivatives which vanish in the limit $<(\sum r_{i})^{2}>\rightarrow\infty$. 

The interaction with the condensate particles is taken to be a gauge
theory with one high external mass characterizing the condensates.
The simplest possibility is an additional, sufficiently strongly coupling
high mass component in known gauge fields. Also possible is a new
SU(3) Technicolor interaction with a sufficiently strong interaction
and a high gluon mass. Such a theory might be more attractive for
reasons not connected to our discussion%
\footnote{They might offer an understanding of the nature of generations. They
could be duality-connected to theoretically attractive brane theories.%
}. The value of the new high mass scale is not intrinsically connected
to the $W$ mass.

To lowest order there are then the two contributions shown in figure
1, where the separation between visible world and cosmological vacuum
is indicated by a box.

\begin{figure}
\includegraphics[%
  width=1.03\textwidth,
  keepaspectratio]{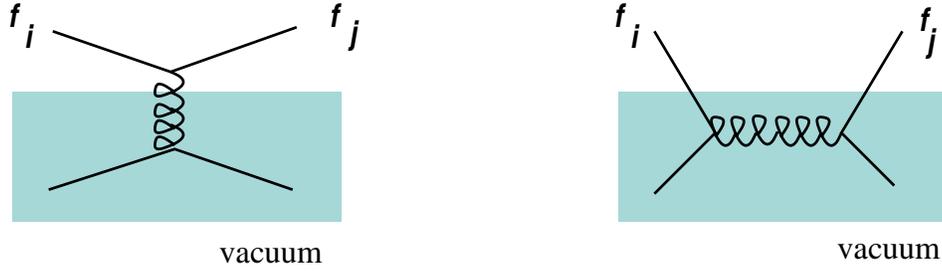}

\caption{Fermion interactions with cosmographic vacuum}
\end{figure}

The first term involves tensors of rank 1 and vanishes in the considered
limit. The Fierz transformation of the second term includes a scalar
contribution which survives. 

The condensate interaction interaction has to be strong and higher
orders have to be included. The high binding energy disallows intermediate
unbound condensates on the corresponding scale. This requires the
fermion exchange process to stay point like. Except for the strength
of the effective coupling nothing changes on a qualitative level.
There is no such argument for a possible two-gauge-boson Lorentz-scalar
contribution to the first term. Such a contribution could play a role
as a seed for the gluon condensate needed for chiral symmetry breaking
not considered here.

The basic result is reassuring. As needed the interaction with the
vacuum turns out to be flavor dependent, with the mass of a fermion
$m_{i}$ proportional to its density $\rho_{i}$ in the cosmological
vacuum. The exact form of this dependence can be estimated in lowest
order to be: $m_{i}\sim\rho_{i}g^{2}/\widetilde{M}^{2}$ where $\widetilde{M}$
is an effective mass of a boson exchanged in the vacuum. 

In the considered zero-momentum limit there is no distinction between
incoming and outgoing particles as indicated in figure 2. %
\begin{figure}
\includegraphics[%
  width=1.03\textwidth,
  keepaspectratio]{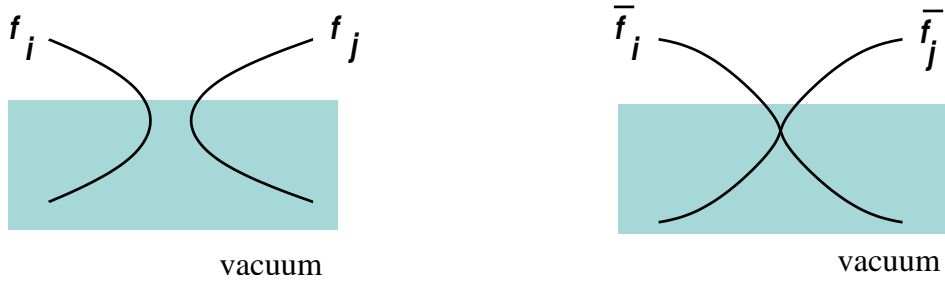}

\caption{The equality of particle and antiparticle masses}
\end{figure}
 Particles and anti-particles have the same mass. In this way CPT
is conserved separately in the visible part of the world.

The magnetic dipol field energy excludes bound states with non scalar
spins from the vacuum. Therefor the cosmographic vacuum cannot serve
as spin reservoir and angular momentum must be conserved in the visible
world. As the scalar nature must hold independently of the Lorentz
system the vacuum spin must vanish separately for each flavor with
it specific mass. Therefore the cosmographic vacuum does not destroy
parity (P) invariance for particles in the visible world.

Not all fermion index combinations $(i,j)$ can exist in the cosmographic
vacuum. To reach the low energy density the vacuum must contain tightly
bound electrically neutral and colorless states. In this way the mass
matrix decomposes into four separate matrices: $M_{u,c,t}$, $M_{d,s,b}$,
$M_{e,\mu,\tau}$, and $M_{\nu(1),\nu(2),\nu(3)}$.%
\footnote{The model contains a conserved fermion number and excludes Majorana
masses. The theoretical motivation for Majorana masses is not strong
as the model will have to accommodate anyhow large differences in
mass ($\frac{m_{t}}{m_{e}}$$\sim10^{5}$). If the neutrino-less double
beta decay is confirmed the vacuum structure assumed in the model
cannot survive.%
}

\begin{floatingfigure}{0.5\columnwidth}%
\includegraphics[%
  bb=0bp 0bp 155bp 191bp,
  width=0.26\textwidth,
  keepaspectratio]{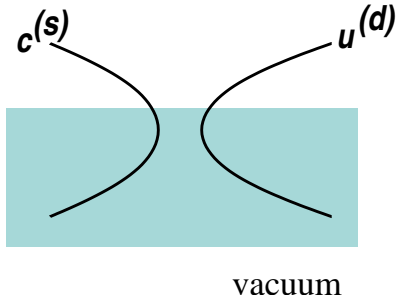}

\caption{Flavor-changing contribution}\end{floatingfigure}%
The result closely resembles the standard approach. Diagonalization
allows to define flavor eigenstates conserved in neutral-current interactions
in the usual way. Charge currents require to consider mass matrices
in a Cabibbo-rotated, non-diagonal basis leading to flavor transitions
in the usual way. However, there are two fundamental differences.

The first point concerns flavor conservation depicted in figure 3.
The superscripts indicate that the transferred basis with the $SU(2)$
partners of the $d$ resp. $s$ quark is considered. All flavors are
conserved if the visible world and the vacuum are taken together!
Only if the visible world is considered separately ``flavor''-changing
processes appear. E.g.\ the matrix element $M(c^{(s)}\to u^{(d)})$
and the corresponding change in the vacuum is responsible  for the
loss of strangeness. The cosmographic vacuum acts as an infinite reservoir. 

Take an $s\overline{s}$ pair produced in $\ $$e^{+}e^{-}$ annihilation
out of the real vacuum. It eventually annihilates from the visible
world ($K$ decays) and leaves a corresponding pair in the cosmographic
vacuum, where it spreads out and where it might eventually annihilate
as part of the cosmologically slow decay of the cosmographic vacuum. 

The second point is more subtle and concerns CP violation. Masses
are a scalar, low momenta limits of amplitudes. As there are no intermediate
states the optical theorem requires these amplitudes - taken by themselves
- to be real. They are not part of a fundamental Hamiltonian and hermiticity
is not a requirement! 

If the cosmographic vacuum contains different $\left<f_{i\,}\overline{f_{j}}\right>$
and $\left<\overline{f_{i\,}}f_{j}\right>$ condensates it accommodates
CP violation in the visible world. Figure 4 lists the possible contribution
of the mass-matrix elements with $s$ and $d$ quarks. Different condensates
lead to a difference between the first and the second line.

\begin{figure}
\includegraphics[%
  width=1\textwidth,
  height=40in,
  keepaspectratio]{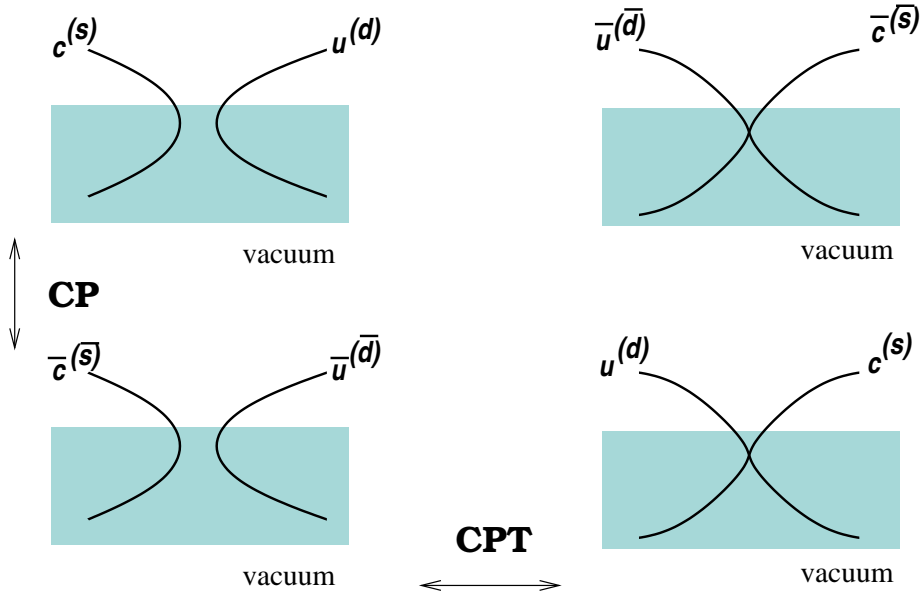}

\caption{The symmetries of the flavor-changing contributions}
\end{figure}

In an asymmetric particle--antiparticle world it is reasonable to
assume that an initial mixture of flavors decayed in the nowadays
seen light flavors. Hence for an asymmetric vacuum slightly different
amounts of p.e. $<s_{\,}\overline{d}>$ than $<d_{\,}\overline{s}>$
could be expected %
\footnote{The simple unitarity requirement would be replaced by a statistical
consideration with the principle of detailed balance.%
}. 

As explained in Sect.~4 in cosmography particle and antiparticle
dominated zones can coexist and a symmetric world can be assumed\cite{Gell-Mann}.
The dominant CP asymmetry left in the vacuum is taken to be just a
slightly different phase of the $\left<f_{i\,}\overline{f_{j}}\right>$
and $\left<\overline{f_{i\,}}f_{j}\right>$ condensates somehow reflecting
the different locations of particle and antiparticle zones. As seen
below this changes the flavor changing $\left(\begin{array}{cc}
0 & 1\\
1 & 0\end{array}\right)$ entry to the usual $\cos\delta\left(\begin{array}{cc}
0 & 1\\
1 & 0\end{array}\right)+\sin\delta\left(\begin{array}{cc}
0 & -i\\
i & 0\end{array}\right)$ contribution. Following the usual hermitian theory $\delta$ can
be obtained from a comparison of the $K^{0}\leftarrow\overline{K^{0}}$
transition with the $K^{0}\rightarrow\overline{K^{0}}$ transition
-- which involves twice the first resp. the ~second line.

The most specific evidence for CP violation comes from interference
experiments. On the first sight the different final vacuum states
seem to completely exclude such contributions. The vacuum part of
the process has to be included in the consideration. The nowadays
vacuum is on the experimental scale constant. There is no phase which
depends on the locations where strangeness is given resp.\ taken
from the vacuum. Also the vacuum is by now coherent in regard to its
bosonic bound states. The phases of all incoming and outgoing bosons
are fixed. In this way the coherence needed for the observed interference
is not destroyed even in a complicated process in which, say, in one
particular contribution two $s$ quarks are taken from two vacuum
bosons at the site of the $K_{{\rm {S}}}^{0}$ and the $K_{{\rm {L}}}^{0}$
decay and two $s$ quark are given to two different vacuum bosons
at the site of, say, a $\Delta S=-2$ self interaction of the $K_{{\rm {L}}}^{0}$.
The vacuum asymmetry in regard to the phase of $s\bar{d}$ and $\bar{s}d$
leeds to a phase difference between a replacement of an $s$ by a
$d$ and again of a $\bar{d}$ by an $\bar{s}$ and a replacement
of an $\bar{s}$ by a $\bar{d}$ and one of a $d$ by an $s$ yielding
the CP violating contribution to the mass matrix described above. 

In contrast to the usual model there can be CP violation with two
generations and the CP violation can be adjusted separately for each
flavor pair. If it is indeed based on the particle--antiparticle
fluctuations in the visible light quark world, it should be less for
$b\bar{c}$ and $t\bar{c}$ systems than for $u\bar{c}$, $u\bar{t}$,
$d\bar{s}$ and $d\bar{b}$ systems. In the standard model this observation
is obtained from an arbitrary parameterization of the unitarity triangle. 

To summarize CP violation arises as one restricts the consideration
to the visible world and ignores the asymmetries of the cosmographic
vacuum. The cosmological need for CP violation to cause particle--antiparticle
asymmetry does not exist in a landscaped universe. 

To be more specific about the possible dynamic: what could such a
vacuum look like? With one important caveat discussed later, the cosmographic
vacuum could consist of very tightly bound neutral, colorless $f_{i\,}\overline{f_{j}}$
pairs. In contrast to top-condensate models\cite{Lindner,Bardeen}
the cosmographic model does not allow for a special flavor selection
in the Lagrangian. As consequence the cosmographic vacuum must be
more complicated than a pure top condensate. Spatial considerations
might be essential and, unpleasantly, the model to a certain degree
lacks predictiveness.

For the condensation three ingredients must be assumed: 

\begin{description}
\item [a)]If a fermion somehow condensates and obtains a heavier mass closer
to the new mass scale, it can condensate more strongly causing an
increased density. The self-propelling mass asymmetry could be an
essential part of the explanation of the large variation in the densities
resp.\ masses. 
\item [b)]The interaction between different condensate states must somehow
prevent other fermions from condensation to similar masses. Packaging
geometry might play a role. The argument is again needed to support
the dominant role of the top quark in the vacuum. The same mechanism
might work to a certain degree for the $b$ $s$ $d$ fermion system
and to lesser an extend for the less dense $\tau$ $\mu$ $e$ and
the $\nu$ fermion systems. 
\item [c)]The hierarchy of the observed masses \cite{Donoghue} should
also reflect fluctuations from the $t\overline{t}$ condensate to
other fermion condensates. The probabilities of such fluctuations
depend on the ``similarity'' of the bound state. Whether the binding
is also supported by colors or electric charges should be the defining
properties for the ``similarity''. In this way the order of the $u$...
, the $d...$, the $e...$and the $\nu...$fermion masses might be
understood.
\end{description}
It is important to keep in mind that these are just general constraints
on a largely accidental cosmological vacuum.

\section{The Vector-Boson Masses}

\begin{figure}
\includegraphics[%
  width=0.9\textwidth,
  keepaspectratio]{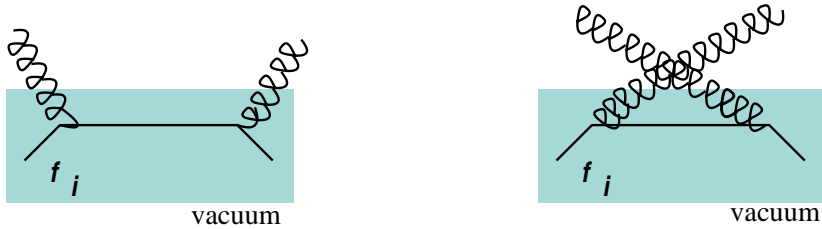}

\caption{Vector-boson interactions with cosmographic vacuum}
\end{figure}
Vector bosons can also interact with fermions of the cosmographic
vacuum in the way shown in figure 5. Obviously both vector bosons
pictured have to have corresponding charges. The electric neutrality
of the vacuum prevents a mass of the $\gamma$ through this mechanism.

In the lowest-order Operator Product Expansion one obtains the term:\[
~\sum_{i}\left(\rho_{i}\theta_{i}\,\vec{W}_{\mu}\vec{W}^{\mu}+\rho_{i}\theta_{i}\, B_{\mu}B^{\mu}\right)\]
where $\rho_{i}$ is the density of the \emph{i}th fermion and where
$\theta_{i}=1$ if the process is allowed for the considered fermion
type. There also can be mixed terms.

The contribution of a single fermion to the squared vector-boson mass
is proportional to the density and in lowest order to the square of
the coupling: $M_{W}^{2}\sim\sum(\rho_{i}g^{2}/\widetilde{m_{i}})$,
where $\widetilde{m_{i}}$ is the effective mass of a fermion in the
cosmographic vacuum.

In spite of the similarity the interactions responsible for fermion-
and vector-boson masses are actually quite different. The fermion
exchange process should depend on the binding like a hadronic interaction
while the vector-boson interactions essentially just count the charges. 

The relative weights of these processes are uncertain.%
\footnote{It is even possible that the fermion condensate is practically unrelated
to the vector-boson masses and conventional Higgs boson fields play
a dominant role for the vector-boson masses while the fermion part
of the condensate could be responsible for fermionic masses.%
} \begin{floatingfigure}{0.5\columnwidth}%
\includegraphics[%
  bb=0bp 0bp 699bp 150bp,
  width=58mm,
  height=27mm]{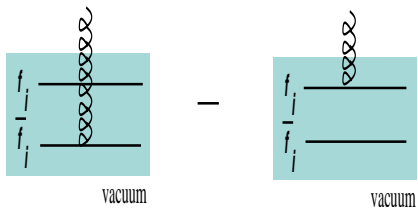}

\caption{Mesons and $B$ mass}

~\end{floatingfigure}%
Here we will consider the possibility that vector-boson masses are
also a consequence of the fermion condensate. Encouraging is the following.
As the top dominates ($m_{t}/m_{b}\sim20$) the vacuum contribution
to the vector bosons there is a simple estimate explaining the observed
similarity of the vector-boson and top-quark mas\-ses.

The identification requires a complication in the vacuum structure.
To understand it, we first need to explain a counting rule. Considering
the contributions of fermions $f$ one finds\begin{floatingfigure}{0.5\columnwidth}%
\includegraphics[%
  bb=-100bp 10bp 383bp 299bp,
  width=40mm,
  height=25in,
  keepaspectratio]{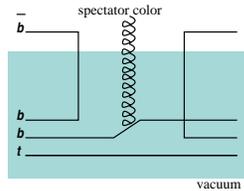}

\caption{Cancellation of the $b$-quark contribution}\end{floatingfigure}%
\begin{eqnarray*}
M_{W}^{2} & \sim & \# f+\#\overline{f}\\
M_{B}^{2} & \sim & \# f-\#\overline{f}\end{eqnarray*}
 where $B$ is the U(1) gauge boson. As indicated in figure 6 mesons
cannot contribute to $M_{B}$. The Weinberg angle $\tan\theta_{{\rm {W}}}=M_{B}^{2}/M_{W}^{2}$
thus determines the fermion or anti-fermion excess in the vacuum.

The excess presents no problem as the condensate can contain colorless,
electrically neutral neutron-like states. Both the enhanced usual
gauge and the Technicolor scenario allow for such mixed symmetry states.
As the vacuum has to be spinless these baryons have to come in pairs
with compensating spins. A mixture of two $(qqq)$ baryons and exactly
seven $(q\overline{q})$ mesons leads to a Weinberg angle of $\tan\theta_{{\rm {W}}}=3/10$
or $\sin\theta_{{\rm {W}}}=0.2308$ pretty much in the center of the
experimental value.

A possible problem arises with the top dominance. A neutral baryon
(like $(tbb)$) contains at most one top quark. Ignoring all other
flavors this would lead to a ratio $m_{t}/m_{b}=4$ instead of $20$.
However, there is a negative interference between both $b$ quarks
and the mixed contribution illustrated in figure 7 could lead to the
required cancellation if even and odd parity color singlet exchanges
are of similar weight.

\section{Experimental Expectations }

The condensate bound states of the cosmographic vacuum are more tightly
bound than ordinary particles. Do \textbf{similar} \textbf{particles}
exist \textbf{in the visible world?}

Three of such states are needed outside of the vacuum for the new
degrees of freedom in the massive vector bosons. They are presumably
flavor mixtures dominated by $t\overline{b}$, $\frac{1}{\sqrt{{2}}}(t\overline{t}-b\overline{b})$,
and $b\overline{t}$ .

Little is known about other Higgs like bosons. It might be necessary
that they or at least the singlet partner $\frac{1}{\sqrt{{2}}}(t\overline{t}+b\overline{b})$
exists with a mass in the range of the vector bosons masses. The singlet
partner would then be similar to the standard model Higgs except there
would be no preference for a $t\overline{t}$ against a kinematically
favored $b\overline{b}$ decay. In general, each Higgs like boson
couples to the fermions it contains. The typical production leads
to a dominance of light fermion decays. 

However, the argument with the $W$ mass range is not very strong
-- dispersion of light is seen at energies decades below atomic binding
energies. The usual masses depend on a combination of condensation
effects and of cosmological scales; the masses of Higgs like particles
should reflect the condensation mass scale only. There is no good
argument to keep the condensation mass in a particular range. The
hierarchy problem between this condensation mass and the unification
mass is outside of the consideration. 

Experimentally no problem is expected for the luminosity needed to
produce such particles: couplings are part of pure physics and this
excludes small numbers. However, the energy will be critical as such
particles are known to be heavy. The absence of abnormal backward
scattering in $e^{+}e^{-}$annihilation at LEP limits the corresponding
``Higgs''-boson to $M(H_{\{ e^{+}e^{-}\}}>189){\rm {GeV}\,}$\cite{Bhabha}.
The large-transverse-momentum jet production at Fermilab limits $M(H_{\{ u\overline{u}\}})$
to an energy above $1{\rm {TeV}\,}$\cite{Fermilab}. At LHC a multi-TeV
range will be reached. 

The cosmographic vacuum would obviously affect many \textbf{cosmological
arguments}. Central is the question of the uniformity of the vacuum.
The best bet is to look for changes in the symmetry breaking of the
vacuum which occurred latest, as later means presumably also closer.
Can we see cosmographic domains with different chiral symmetry breaking?

The first step in understanding this question is a modified \textsc{Leibnitz
theodicy} stating that we live in ``\textit{one of the most visible
of all cosmographic domains}''. Slight variations in the chiral symmetry
breaking, which affect nucleon masses but not lepton masses, would
prevent fusion or allow for a neutron- or hydrogen-dominated world. 

So if variations exist we will not see them. Invisible domains might
just contribute to the unseen mass of the universe. Starless domains
can act as a buffer to allow for a not visibly violent coexistence
of particle and antiparticle dominated zones. The asymmetry could
be a condensation effect. The freeze out time in zones with non-vanishing
baryonic quantum numbers might be earlier \cite{Zhitnitsky}.

Slight variations in spectroscopic measurements might offer an exception
to invisibility. Another most visible region could have a still admissible
shift in chiral symmetry breaking with the corresponding shift in
the nuclear radius. An appealing idea is that the claimed $10^{-5}$
variation in the fine structure constant with \textit{time} seen in
heavier atoms and not in lighter ones \cite{Spectrum} could possibly
be re-analyzed and re-interpreted as an \emph{zonal} effect of a $10^{-2}$
change in the nuclear mass and radius in a still not invisible domain
entering the hyperfine structure of the spectrum as $E_{hyperfine}\propto m_{e}/m_{p}E_{fine}$.
It is possible that the first evidence for the concept of a non-universal
vacuum structure can be obtained in such a way.

\section{Conclusion }

The paper attempts actual explanations and not just parameterizations
of observations. Many of the presented ideas are quite imaginative
and require more scrutiny. One purpose of this paper is to illustrate
that mass generation is still largely unsettled. If Higgs-like bosons
are produced at LHC they might not have a preference to decay in heavy
flavors.

\end{document}